\documentclass[prb,twocolumn]{revtex4}
\usepackage{graphicx} %
\usepackage{bm}       
\usepackage{dcolumn}  
\usepackage{amsmath,amssymb}

\newcommand{\bco}{\bm{c}}
\newcommand{\bk}{\bm{k}}
\newcommand{\bA}{\bm{A}}

\newcommand{\bPsi}{\bm{\psi}}

\newcommand{\eas}[0]{\begin{eqnarray*}}
\newcommand{\eae}[0]{\end{eqnarray*}}
\newcommand{\les}[0]{\begin{equation}}
\newcommand{\lee}[0]{\end{equation}}
\newcommand{\leas}[0]{\begin{eqnarray}}
\newcommand{\leae}[0]{\end{eqnarray}}

\newcommand{\mchss}[4]
{
\left\{
\begin{array}{cc}
#1 & #2   \\
#3 & #4
\end{array}
\right.
}

\newcommand{\mat}[4]
{
\left(
\begin{array}{cc}
#1 & #2 \\
#3 & #4 
\end{array}
\right)
}
\newcommand{\mvec}[2]
{
\left(
\begin{array}{c}
#1  \\
#2  
\end{array}
\right)
}

\begin{document}


\title{
Bulk-edge correspondence in graphene
 with/without magnetic field:\\
Chiral symmetry, Dirac fermions and Edge states
}

\author{Y. Hatsugai}
\email{hatsugai@sakura.cc.tsukuba.ac.jp}
\affiliation{
  Institute of Physics, 
University of Tsukuba, Tsukuba, Ibaraki, 305-8571, Japan }

\date{\today}

\begin{abstract} 
There are two types of edge states in graphene with/without
magnetic field. One is 
a quantum Hall edge state, which is topologically
protected against small perturbation. 
The other is a chiral zero mode that is localized near the 
boundary  with/without magnetic field.
The latter is also topological but is guaranteed to be zero energy
by the chiral symmetry, which is also responsible for 
massless Dirac like dispersion.
Conceptual roles of the edge states 
are stressed and reviewed  from a view point of
the bulk-edge correspondence and the topological order. 
\end{abstract}


\maketitle


\section{Introduction}
\vskip -0.3cm
Many of  physical states of matter  are 
characterized by order parameters
based on the symmetry breaking.
Especially a symmetry with continuous parameters
has special importance in a quantum state of matter. 
When the ground state of the system with continuous 
symmetry does not posses the symmetry, that is, the ground 
state is not invariant against the symmetry operation, 
there exists a gapless excitation as the Nambu-Goldston mode.
A typical  example is the Heisenberg magnet
with antiferromagnetic order ( Neel order ). 
The ground state of the spherical symmetric ( in spin space )
hamiltonian is not invariant against the spin rotation. 
Then above the ground state, there is a gapless spin excitation
as the spin wave.
This excitation is realized by  infinitesimally small spin deformation
over infinitely 
wide range as was done by the Lieb-Schultz-Mattis\cite{LSM}. 

Although it has been quite successful to characterize many of phases, 
it has been realized that 
there still exist many important physical phases
that are not well described by this spontaneous 
symmetry breaking recipe.
This class of matter includes
many of 
quantum Hall states, 
Haldane spin chains (integer spin chains), 
Kondo insulators and much more. 
A ground state of graphene also belongs to this new class.
 In a ground state with strong quantum fluctuation, 
order formation is strongly suppressed and a quantum ground 
state without any fundamental symmetry breaking is realized. 
In such a case, there is no reason to expect 
gapless excitations, which results in a gapped quantum state.
Phases of this class of matter are
the quantum liquids. Then to characterize such a quantum liquid
phase is one of the important problems.
A novel concept as the topological order 
that was first proposed to discuss the quantum Hall state\cite{Wen89}
is now under active studies for the purpose. 

Due to be the absence of symmetry breaking, 
one needs to use something new to describe the topological order. 
One of such possibilities to use geometrical phases
of the quantum states\cite{Hatsugai04e,Hatsugai05-char}.
It works well theoretically to describe some 
of quantum liquids and spin liquids\cite{Hatsugai06-QBerry,Hirano08}.   
Another tool is to use edge states, which 
are one of the feature of the topological ordered states.
There exist non trivial edge states, as the 
bound states near the boundaries,
 appear when the 
bulk are topologically non-trivial.
Even if the quantum liquid does not posses any symmetry breaking and
featureless, there exist characteristic boundary states when the system 
has  boundaries and impurities. 
These generic edge states are characteristic to the topological ordered 
states and they themselves characterize the quantum states, which we 
call the bulk-edge correspondence\cite{Hatsugai93b}. 
These kinds of non-trivial edge states appear in many different 
ares in condensed matter physics, such as optical lattices\cite{Scarola07}
and photonic crystals\cite{Wang08}.
Further an entanglement entropy,
 which is also useful to characterize the quantum liquids,   
is also directly related to the edge states\cite{Ryu06,Katsura07}.
 In this short review, we have focused on 
non trivial edge states of graphene based on the bulk-edge correspondence
and try to make clear topological aspects of graphene with and without magnetic
field. 

\section{Quantum Hall effect of Graphene}
\subsection{Edge states of the Quantum Hall states}
\begin{figure}
  \begin{center}
  \includegraphics[width=4cm]{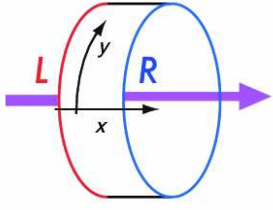}
  \end{center}
 \caption{Cylinder used in the Laughlin's argument. 
Edge states in this review 
are all discussed in this cylindrical geometry. }
  \label{fig:cyl}
\end{figure}

\vskip -0.3cm
As is well known, 
the  Hall conductance of two dimensional electrons has an 
intrinsic topological meaning that is an origin of
high accuracy of the quantization.
This quantization is clearly demonstrated 
by Laughlin using an adiabatic process
with gauge invariance where 
Aharonov-Bohm(AB) flux $\Phi$  passes through the hole of
the cylinder\cite{Laughlin81} (Fig.\ref{fig:cyl}). 
Quantized value of the Hall conductance (in a suitable unit) is 
a number of electrons carried across the system 
when the AB flux is adiabatically increased by 
one flux quantum. 
Considering a system with edges, this quantized but ambiguous integer 
is uniquely specified by the number of edge modes $I_j$ in the $j$-th
Landau gap where 
the Fermi energy lies\cite{Halperin82,Hatsugai93a}
\begin{alignat*}{1} 
\sigma^{\rm Edge}_{xy} =& \frac {e^2}{h} I_j.
\end{alignat*} 
\vskip -0.2cm
It implies that the edge states has essential importance 
for the physics of the quantum Hall effects. 

Although the Hall current is dissipation less, it is natural to 
describe the Hall conductance as a bulk property.
It is given within the linear response theory
as \cite{Thouless82},
$
\sigma^{\rm Bulk}_{xy} =
 \frac {e^2}{h} \sum_{\ell}C_\ell
$ 
where the summation is over the filled Landau levels 
and $C_\ell$ is the first Chern number of the $\ell$-th Landau band,
$
C_\ell = \frac {1}{2\pi i } \int d^2k\,  ({\rm rot}_k\,  \bA_\ell )_z
$, $\bA_\ell
  =\langle \psi_\ell| \bm{\nabla}_k \psi_\ell \rangle $, 
where $|\psi_\ell \rangle $
is a one body wave function of 
the $\ell$-th Landau band\cite{Thouless82}.
\begin{figure}
  \begin{center}
  \includegraphics[width=8cm]{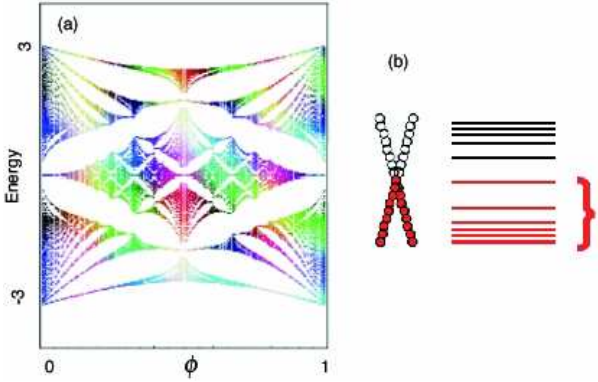}
  \end{center}
 \caption{(a)One particle energy spectrum of graphene under a magnetic field. 
Horizontal axis $\phi$ is a magnetic flux  per hexagon in a unit of 
magnetic flux quantum.
  (b) Schematic Dirac sea without magnetic field and
Landau level of the Dirac fermions
}
  \label{fig:hof}
\end{figure}

\begin{figure}
  \begin{center}
  \includegraphics[width=6cm]{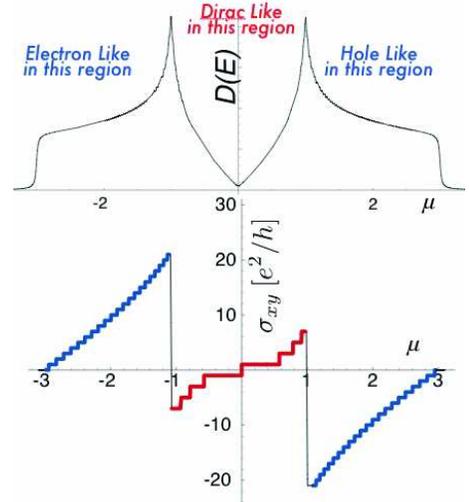}
  \end{center}
 \caption{Hall conductance of the graphene as a function of
the chemical potential $\mu $. ( $\phi=1/31$)}
  \label{fig:dossxy}
\end{figure}

One particle spectrum of the graphene in a magnetic field is 
given in Fig.\ref{fig:hof}. 
When the Fermi energy is set around the 
$E=0$, one needs to 
fill the Dirac sea, 
which causes numerical difficulties.
In such a case, following non Abelian formulation for the Hall conductance is
useful 
\begin{alignat}{1} 
\sigma_{xy}^{\rm Bulk} =& \frac {e^2}{h} C_F,\ \
C_F = \frac {1}{2\pi i} \int d {\rm Tr \,} \bA\label{eq:chern}\\
\bA =& \bPsi ^\dagger d \bPsi = 
\bPsi ^\dagger \partial _\mu \bPsi  d x^\mu
\nonumber \\
\bPsi =& (|\psi_1 \rangle ,\cdots,|\psi_j \rangle )
\nonumber 
\end{alignat} 
\vskip -0.2cm
where $ (x^1,x^2)=(k_x,k_y)$\cite{Hatsugai04e,Hatsugai05-char,Hatsugai06PRB}.
Especially as for the numerical evaluation 
of the weak field limit.
By using this method
with a  technique invented in the lattice gauge theory
\cite{fukui2005},
the Hall 
conductance of the graphene as a function of the chemical potential $\mu$
is calculated without any numerical difficulties\cite{Hatsugai06PRB}.
By this technical but important development, 
we have further evaluate the Hall conductance of realistic graphene
with multi-band orbitals\cite{Arai08}.
The results by the single band case is 
plotted in Fig.\ref{fig:dossxy} with the density of states (DOS)
without  magnetic field\cite{Hatsugai06PRB}.
Three different behaviors ( electron like,
 Dirac like and hole like ) are clearly observed 
which are separated by the singularities of
the DOS.
Near $E=0$, it reads 
$ 
\sigma_{xy}^{\rm Bulk } =
 \frac{e^2}h (2N+1)
$ with integer $N$. This is a Hall conductance of the Dirac fermions.

\begin{figure}
  \begin{center}
  \includegraphics[width=7cm]{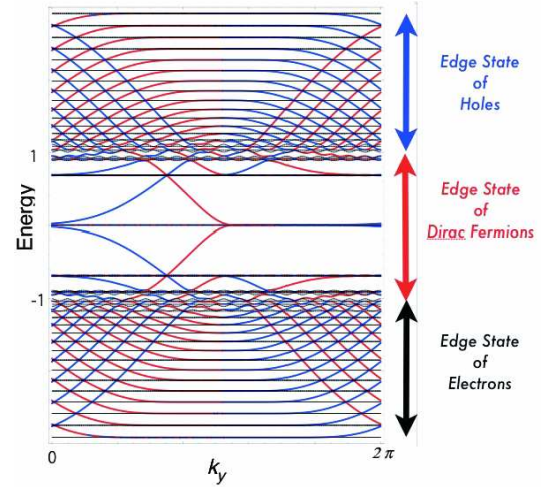}
  \end{center}
 \caption{Energy spectrum of the graphene on a cylinder with zigzag edge
via momentum along the cylindrical direction $k_y$. 
Red and blue lines are energies of the edge states localized at right 
and left edges. ($\phi=1/21$)}
  \label{fig:edge}
\end{figure}

Although he Laughlin's argument is enough to guarantee
the integral nature of the
Hall conductance, only the generic consideration can not fix this 
integer. 
It is specified by considering edge states that is implicitly 
assumed in the Laghlin argument\cite{Hatsugai93a}.
The Hall conductance  is given by the number of edge modes 
in the energy gap that the Fermi energy lies.
In Fig.\ref{fig:edge}, energy spectrum of graphene on the cylinder
( with zigzag boundary ) is shown. The red and blue lines in the energy
gap regions are the energies of the edge states localized near the 
right and left boundaries. 
Counting the number of edge mode in the energy
gap region, 
we have obtained the Hall conductance of the 
graphene with edges
which does coincide to that
 given by the Chern number, eq.(\ref{eq:chern}).
$
\sigma _{xy } ^{\rm Bulk} = 
\sigma _{xy } ^{\rm Edge}
$.

\subsection{Bulk - Edge correspondence}
\begin{figure}
  \begin{center}
  \includegraphics[width=8cm]{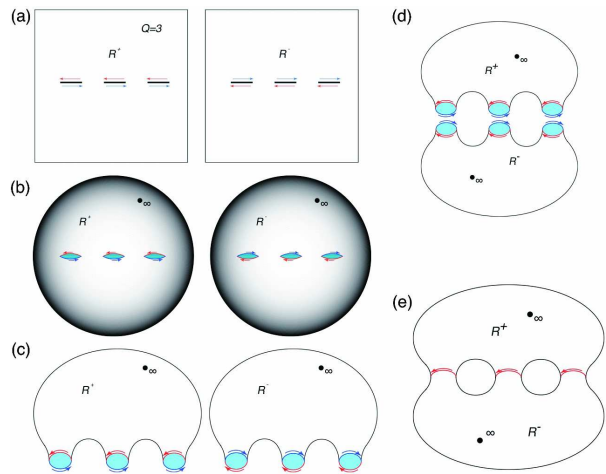}
  \end{center}
 \caption{Construction of the Riemann surface of the Harper equation}
  \label{fig:RS}
\end{figure}

\vskip -0.3cm
It can be natural to expect
that  two expressions for the Hall conductance by the bulk and by the edges 
give the same value
 since both describe the same physics. 
It is actually proved even mathematically as\cite{Hatsugai93b,yh-qhe-review}
\begin{alignat*}{1} 
\sigma^{\rm Edge}_{xy} =&\sigma^{\rm Bulk}_{xy} 
\end{alignat*} 
which is also written as 
\vskip -0.4cm
\begin{alignat}{1} 
C_j =& I_j-I_{j-1}.
\label{eq:bulkedge}
\end{alignat} 
\vskip -0.2cm
This general structure is also applied for the graphene\cite{Hatsugai06PRB}. 
As discussed in the previous section, the Chern number $C_j$ is
a topological quantity. 
Also the number of edge modes, $I_j$ has also topological 
meaning.
Here let us give rough idea of  the topological meaning 
of the edge modes\cite{Hatsugai93a}. 
When one discusses  usual electrons in a magnetic field $B$
 in a Landau gauge in $x$-direction, two dimensional hamiltonian is decomposed 
into the sum of the one dimensional hamiltonian of the harmonic oscillators
(with parameter $k_y$). 
When one describes the Bloch electrons as graphene on the 
cylinder with the Landau gauge, 
similar decomposition of the two dimensional hamiltonian is 
possible. In this case, corresponding one dimensional system with
parameter $k_y$ is not a simple harmonic oscillator but
the Harper equation\cite{yh-qhe-review,Hatsugai93a}.
\vskip -0.4cm
\begin{alignat*}{1} 
H&(\text{2D electrons with } B) = \sum_{k_y}H_{k_y}(\text{harmonic oscillator})
\\
H&(\text{graphene with } B) = \sum_{k_y}H_{k_y}(\text{harper problem})
\end{alignat*} 
\vskip -0.4cm
To describe the edge states ( bound states ) and the Bloch 
states ( scattering states ) on the same footing, 
we need to consider a complex energy surface.
As for the Harper equation, there are multiple ($q$) energy bands that
 correspond to the Landau levels.
We need to prepare two complex energy planes $R^+$ and $R^-$
by making these $q$ energy bands
as branch cuts as in Fig.\ref{fig:RS}(a).  To discuss
the problem, it is useful to make them into Riemann spheres
by identifying the infinities into each single points.
Then the two Riemann spheres are glued into one 
along the branch cuts (Fig.\ref{fig:RS}(a)-(e).) In Figs.\ref{fig:RS},
we have shown this procedure symbolically for $q=3$ case. 
Finally we have Riemann surface (RS) with $g=q-1$ holes ( handles ), which 
coincide to the energy gaps of the Harper equation. 
On this RS, generic  Bloch function has several zeros which correspond 
to the energies of the edge states. One can identify the edge of the
cylinder where the edge states localize by the position of the zeros.
When it is on the $R^+$ ($R^-$), the edge 
state is localized at the right (left) edge\cite{footnote}.
This is for fixed $k_y$. Then by changing $k_y$, these zeros 
of the Bloch state ( edge state energies ) move and form loops
since $k_y=0$ and $k_y=2\pi$ to be identified. 
Then  the winding numbers of
the
loops around the holes ( gaps ) of the RS give the numbers of edge 
state, $I_j$ 
is defined.
By using the Laughlin's argument, the winding number $I_j$ gives
the Hall conductance 
when the chemical potential lies in the $j$-the gap\cite{Hatsugai93a}.
In this way, the 
relation eq.(\ref{eq:bulkedge}) implies a bulk-edge correspondence
of the topological quantities\cite{Hatsugai93b}. 
It is clearly demonstrated 
in the graphene quantum Hall effects\cite{Hatsugai06PRB}.

\section{Chiral Symmetry and zero mode Edge states}
\subsection{Universality of the Dirac fermions and the chiral symmetry}
\vskip -0.3cm
As discussed in the previous section, graphene under a magnetic field
has edge states, which are topologically stable and characteristic
to the quantum Hall effects.
Since the graphene has a chiral symmetry, there are 
additional edge states,
which are  protected not only by the 
topological constraint but also by the chiral symmetry. 
The chiral symmetry requires the edge states to have the 
zero energy as discussed below.

\begin{figure}
  \begin{center}
  \includegraphics[width=8cm]{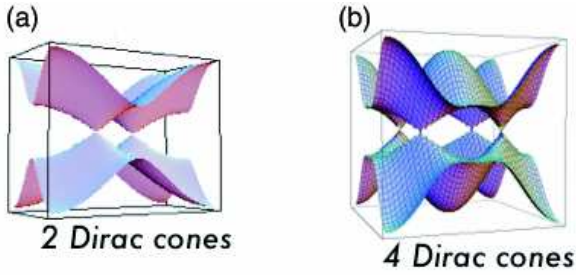}
  \end{center}
 \caption{
Typical energy dispersions for (a) graphene  and (b) two 
dimensional  $d_{x^2-y^2}$-wave 
superconductor.
}
  \label{fig:dirac24}
\end{figure}

As is well known the energy dispersion of graphene is 
the Dirac fermions like and vanishing linearly at some momentum 
(Fig.\ref{fig:dirac24}(a)).
Another example of the  Dirac like dispersion in two dimensions 
is also known in condensed matter. That is the  $d_{x^2-y^2}$-wave 
superconductor 
 (Fig.\ref{fig:dirac24}(b))
which has been discussed 
for the high-$T_C$ materials\cite{Lee93,Hatsugai93c,Ryu02}.
In momentum representation, 
the hamiltonians for both cases are
written as
\begin{alignat*}{1} 
{\cal H} =& \sum_k \bco_k ^\dagger  H_k \bco_k
\end{alignat*} 
where $\bco_k$ is a two component spinor of the fermion operators. 
They are respectively given as $\bco_k=\bco_k^g$ and $ \bco_k=\bco_k^s$, 
$H_k=H_k^g$ and $H_k=H_k^s$ for the graphene and the  $d_{x^2-y^2}$-wave 
superconductor respectively 
where
\begin{alignat*}{1}
H_k^g =& 
\mat
{0}{D(\bk)}
{D^*(\bk)}{0},
\quad 
\bco_k^g =\mvec{c_{\circ} (\bk)}{c_{\bullet}(\bk)}
\\
H_k^g =& 
\mat
{\epsilon(\bk) }{\Delta (\bk)}
{\Delta^*(\bk)}{-\epsilon(\bk) },
\quad 
\bco_k^s =\mvec{c_{\uparrow} (\bk)}{c_{\downarrow} ^\dagger (\bk)}
\end{alignat*} 
As for the graphene, we have assumed that the hopping is only allowed 
between the different sublattices $\circ $ and $\bullet$.
Then the one particle hamiltonian has a chiral invariance

\begin{alignat*}{1}
\{H_k^g,\Gamma^g\}=0, \quad \Gamma^g=\mat{1}{0}{0}{-1}, (\Gamma^g)^2=1
\end{alignat*}

As for the  $d_{x^2-y^2}$-wave superconductivity, when the order parameter does 
preserve the time reversal, the order parameter, $\Delta(\bk)$,
is real
($\Delta(\bk)\in \mathbb{R}$).
In this case, it also has a chiral invariance 
\begin{alignat*}{1}
\{H_k^s,\Gamma^s\}=0, \quad \Gamma^s=\mat{0}{1}{-1}{0}, (\Gamma^s)^2=1
\end{alignat*}
Then the energy dispersion is given by
\begin{alignat*}{1}
E(\bk) =& \pm |Z(\bk)|
\\
\mathbb{C}\ni Z(\bk) =& 
\mchss
{D(\bk)}{\text{ for the graphene}}
{\epsilon(\bk)+ i \Delta (\bk) }{\text{ for the TR invariant
 $d$-wave}}
\end{alignat*} 
It guarantees existence of even number of Dirac fermions, 
since the zero of the $Z(\bk)$ gives the linearly vanishing 
energy gap (generically) and 
the closed curve $C(k_y)=\{Z(k_x,k_y)|k_x\in [0,2\pi]\}$
 moves on the complex $Z$ 
 plane under the condition $C(0)=C(2\pi)$\cite{Hatsugai07EPJ}. 
It implies the curve cut the origin in the complex plane even number of
times when one change $k_y:0\to 2\pi$.

From this two examples, we have realized that there 
are universality of the Dirac like dispersion when the 
system has a chiral symmetry.
This universal feature further guarantees the physics of 
boundary states as discussed below.

\subsection{ Berry phases and the Edge states}
\vskip -0.3cm
The doubling of the generic
Dirac fermion with chiral symmetry is a universal property of the bulk.
This bulk property also restricts the existence of special edge states
as an example of the bulk-edge correspondence. 
As for the 
chiral invariant system, generic condition for the
existence of the zero mode on a cylinder (Fig.\ref{fig:cyl}) 
are given\cite{Ryu02,Ryu_prv}. 
When one takes a momentum along the cylinder as $k_y$, 
the Berry phase ( Zak phase ) for each $k_y$, $\gamma (k_y)$, 
of the one dimensional hamiltonian $H_k$ ($k_y$ fixed) 
is defined  using the Bloch state $|\psi_k \rangle $,
($H_k|\psi_k \rangle =  E(\bk )|\psi_k \rangle $),
\begin{alignat*}{1}
i \gamma (k_y) =& \int_0^{2\pi} dk_x\, A(\bk),\ \
A(\bk) = \langle \psi_k|\frac {\partial  }{\partial k_x  } |\psi_k \rangle 
\end{alignat*} 
Of course, it does have an ambiguity of the gauge ( phase of the 
Bloch state $|\psi_k \rangle $,
$\gamma(k_y)$ is well defined up to modulo $2\pi$.\cite{Hatsugai06-QBerry}.
Generically speaking the Berry phase can takes any values. However 
 the chiral symmetry of the hamiltonian requires the Berry phase 
has to be quantized into $0$ or $\pi$.
Using this quantized 
Berry phases of the bulk,
one can  {\em guarantee the existence of the edge states } 
for an infinitely long (along $x$-direction) cylindrical system when 
the Berry phase  $\gamma(k_y)$ is $\pi$ ($\text{mod}\ 2\pi$)
as far as the {\em
 chiral symmetry is still preserved with edges}\cite{Ryu02}.
\begin{alignat*}{1}
& \quad\quad \gamma (k_y) = \pi\quad \text{mod}\, 2\pi  
\Rightarrow  \\
& \text{Two zero energy localized modes }
\\
&\quad\quad 
\text{near the
at the right and left boundaries of the cylinder}
\end{alignat*} 
 To carry out concrete evaluation of the Berry phases analytically,
  one needs to fix the gauge\cite{Hatsugai04e}. 
However, this gauge fixing is not necessary, at least, to
obtain numerical value\cite{Hatsugai06-QBerry}.

\begin{figure}
  \begin{center}
  \includegraphics[width=9cm]{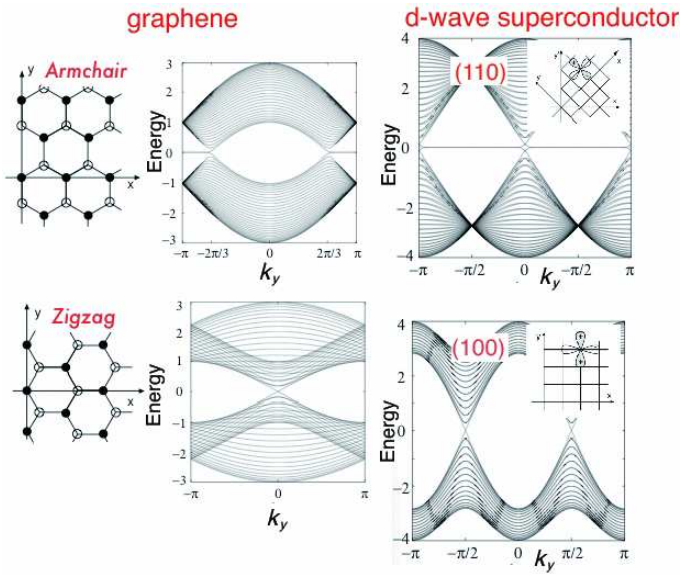}
  \end{center}
  \caption{ Energy spectrum of the graphene and $d_{x^2-y^2}$-wave
    superconductors with different angles on cylinders.  }
  \label{fig:edgeboth}
\end{figure}

Since this condition is applied for each momentum  $k_y$, 
the zero energy edge states form doubly degenerate flat bands
when one considers the original two dimensional problem on the cylinder
(Fig.\ref{fig:cyl}). 
Generically speaking, this condition is not satisfied for  all $k_y$, 
then the flat bands only exist for restricted momenta. 
Applying this generic consideration for graphene, one predicts 
flat band zero energy edge states for $1/3$ of the total momentum 
along $y$ direction near the zigzag edge (Fig.\ref{fig:edgeboth})\cite{Ryu02},
 which was found 
by fujita et al.\cite{Fuji96}. 
This edge mode does not exist for armchair edges. It is also 
consistent with this generic condition (Fig.\ref{fig:edgeboth}).

As for the  $d_{x^2-y^2}$-wave superconductor,
we have applied it and obtain the flat band zero mode edge states
for (110)-direction\cite{Ryu02}.
As for the (100)-direction, we predict 
absence of the zero energy edge states\cite{Ryu02} ( There is no 
topological reason to have the edge states.   See Fig.\ref{fig:edgeboth}).
This is consistent with the 
existence of the Andreev bound states only for the (110)-direction,
which 
has been observed 
in the zero bias conductance measurements\cite{Hu94}.

\subsection{Spontaneous local chiral symmetry breaking}
\vskip -0.3cm
As discussed, graphene and the  $d_{x^2-y^2}$-wave superconductors
have  zero energy edge states protected by the chiral symmetry.
It is topological in the sense that the 
quantized Berry phase ( Zak phase ) is a key (bulk) quantities to 
guarantee the edge states. 

These zero energy edge bands are completely flat ( $E=0$ )
if they exist. The flatness is guaranteed by the chiral symmetry.
Of course, the symmetry of the bulk can not be changed by the
existence of the boundaries.  However, edge potential induced by 
the edges may break the bulk symmetry. This symmetry breaking is
allowed to exist only near the boundaries (by the edge potential).
When one regards the flat band zero modes as a one-dimensional system,
it is natural to expect the Peierls instability that reduces the symmetry
since the flat bands have a diverging density of states. 
Then what would be expected is that the chiral symmetry has broken only
near the edges to make the flat band dispersive or making a gap 
in the edge mode bands. This spontaneous symmetry reduction 
occurs only local near the boundaries\cite{Ryu04}. 
This actually occurs both in the graphene and the  
$d_{x^2-y^2}$-wave superconductors.
In the case of graphene, it corresponds to the 
 boundary magnetic moments near the zigzag edges.
When one considers this boundary magnetism within the mean field theory, 
local moments induce site dependent potentials. It destroys
the chiral symmetry of the graphene near the boundaries.  
It is confirmed, at least, numerically
 based on the density functional theory
calculation\cite{Okada01}.

The same local symmetry breaking also occurs in the
 $d_{x^2-y^2}$-wave superconductors\cite{Ryu03cb}.
In this case, the chiral operator corresponds to the
time-reversal operation that implies real order parameters of
the superconductors.
Then what happens is that the order parameter of the
superconductivity becomes complex only near the boundaries.
It implies  spontaneous generation of local fluxes. 
Topologically, the boundaries of the cylinder
and  point like impurities 
are equivalent. Therefore, 
the generic consideration predict
that spontaneous local flux generation near the boundaries
or impurities\cite{Matsu95,Sigrist95}.

\subsection{Chiral zero modes under the magnetic field}
\vskip -0.3cm
\begin{figure}
  \begin{center}
  \includegraphics[width=9cm]{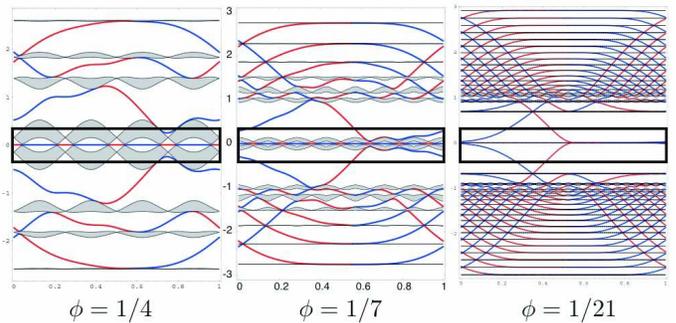}
  \end{center}
 \caption{Energy spectrum of graphene
under  magnetic field
on the cylinder with zigzag edges.
The flux per hexagon is $\phi=1/4, 1/7$ and $1/21$ in  unit of flux quantum
respectively. Red and blue lines 
are the edge states localized near the left and
right edges respectively. The gray regions are the Landau levels.
}
  \label{fig:cedgeB}
\end{figure}
When one considers the graphene under magnetic field, there
are two different type of edge states. The one is 
topologically protected quantum Hall edge states.
The other is chiral zero mode edge states,
 which are topological, but are protected by the chiral symmetry. 
The former is discussed in the previous section. Then 
we discuss the latter one here. 

\begin{figure}
  \begin{center}
  \includegraphics[width=9cm]{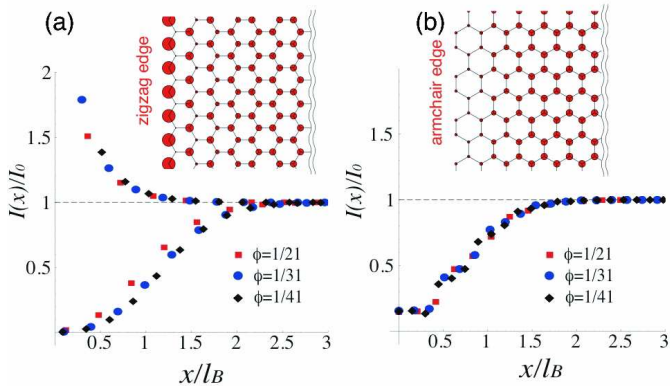}
  \end{center}
 \caption{Local density of states of the graphene with several edges 
around $E=0$. Distance from the edge is denoted by $x$ and $\ell_B$ is a
magnetic length. (a):zigzag edge and (b) armchair edge.}
  \label{fig:cewB}
\end{figure}

As shown in Fig.\ref{fig:cedgeB}, 
there co-exist Dirac fermions' $n=0$ Landau level and the chiral zero mode
edge states for graphene with zigzag edges.
Without magnetic field, there also exist chiral edge modes. However, 
it is degenerate with the bulk Dirac fermion at the gap closing momentum. 
Therefore there is no length scale
for the $E=0$ states.
With the magnetic field, the situation is different and the length scale 
should be determined by the magnetic length scale.
This additional contribution of the boundary state to the local charge
occurs 
only near the zigzag edge. It is absent at the armchair edges 
(See Fig.\ref{fig:cewB})\cite{Arikawa08a}.
Generically speaking, the edge potential repels the charge density
to make depletion layer near the boundary\cite{Halperin82}.
Physically the length scale of 
the depletion layer is determined by the electro-static
potential of the edge detail. Without including it, the length scale 
is the magnetic one $\ell_B$. It is realized at the armchair
edge.(See \ref{fig:cewB}(b)) 

The situation is different at the zigzag edge, there occurs enhancement 
of the local charge density instead of the depletion as shown 
in Fig\ref{fig:cedgeB}(a). 
This is consistent with the existence of chiral zero modes,
which also exist without magnetic field.
However with magnetic field, there also exists
a $n=0$ Landau level of the Dirac fermion at $E=0$.
It plays fundamental role in  reconstruction of the boundary charge. 
Local charge density of this $E=0$ states,
 which has two physical origins
as the chiral edge states and the $n=0$  Landau level,
 has been investigated
by Arikawa et al.\cite{Arikawa08a,Arikawa08b} 
 They have calculated
the local density of state.
It can be
 experimentally observable by the scanning tunneling microscope (STM).
Although there exist  chiral zero modes at the zigzag
edge,
it is not enough to describe the edge charge reconstruction.
The chiral zero modes actually 
show detailed structure
and long tail into the bulk.
However, what is realized in the total local charge 
density is exponentially localised structure
and it becomes to the bulk value
quite rapidly. 
It will be understood as the topological compensation (screening) of
the chiral edge charge by the bulk.
This is a new feature of the boundary reconstruction of the 
$E=0$ states of graphene.

We thank H. Aoki, T. Fukui, S. Ryu, 
M. Arikawa, M. Arai, T. Kawarabayashi and  I. Maruyama for useful 
discussions.
The work is supported in part by Grants-in-Aid for Scientific Research,
No. 20340098, No. 20654034 from JSPS and
No. 220029004 (physics of new quantum phases in super clean materials)
and 20046002 (Novel States of Matter Induced by Frustration) 
on Priority Areas from MEXT.



\begin{thebibliography}{32}
\bibitem{footnote}
As for the continuous 
Schrodinger equation $-\frac {\hbar^2}{2m}\psi''+V\psi=E\psi $,
a formal solution in a constant potential $V$
as $\psi(x)=e^{\pm i x\sqrt{2m(E-V)}/\hbar}$ 
plays an important role by making a branch cut $0\to -i\infty $ in the 
complex plane. Topologically it implies two Riemann spheres into one by gluing them along the branch cuts.
\end{thebibliography}



\end{document}